\documentclass[conference]{IEEEtran}
\IEEEoverridecommandlockouts
\usepackage{amsmath,amsfonts}
\usepackage[mathcal]{eucal}
\usepackage{esint}
\usepackage{algorithmic}
\usepackage{algorithm}
\usepackage{array}
\usepackage[caption=false,font=normalsize,labelfont=sf,textfont=sf]{subfig}
\usepackage{textcomp}
\usepackage{stfloats}
\usepackage{multirow}
\usepackage{booktabs}
\usepackage{etoolbox}
\makeatletter
\patchcmd{\@makecaption}
  {\scshape}
  {}
  {}
  {}
\makeatletter
\patchcmd{\@makecaption}
  {\\}
  {.\ }
  {}
  {}
\makeatother

\usepackage{url}
\usepackage{verbatim}
\usepackage{graphicx}
\usepackage{cite}
\def\BibTeX{{\rm B\kern-.05em{\sc i\kern-.025em b}\kern-.08em
    T\kern-.1667em\lower.7ex\hbox{E}\kern-.125emX}}

\begin{document}

\title{Wireless Communications in Cavity: A Reconfigurable Boundary Modulation based Approach\\
\thanks{National Foundation (NSFC), NO.12141107 supports this work.}}

\author{\IEEEauthorblockN{Xuehui Dong, Xiang Ren, Bokai Lai, Rujing Xiong, Tiebin Mi, Robert Caiming Qiu}
\IEEEauthorblockA{ School of Electronic Information and Communications, Huazhong University of Science and Technology\\Wuhan 430074, China\\
Email:\{xuehuidong, xiangren, bokailai, rujing, mitiebin, caiming\}@hust.edu.cn}
}

\markboth{Journal of \LaTeX\ Class Files,~Vol.~14, No.~8, August~2015}%
{Shell \MakeLowercase{\textit{et al.}}: Bare Demo of IEEEtran.cls for IEEE Journals}

\maketitle

\begin{abstract}
This paper explores the potential wireless communication applications of Reconfigurable Intelligent Surfaces (RIS) in reverberant wave propagation environments. Unlike in free space, we utilize the sensitivity to boundaries of the enclosed electromagnetic (EM) field and the equivalent perturbation of RISs. For the first time, we introduce the framework of reconfigurable boundary modulation in the cavities . We have proposed a robust boundary modulation scheme that exploits the continuity of object motion and the mutation of the codebook switch, which achieves pulse position modulation (PPM) by RIS-generated equivalent pulses for wireless communication in cavities. This approach achieves around 2 Mbps bit rate in the prototype and demonstrates strong resistance to channel's frequency selectivity resulting in an extremely low bit error rate (BER).

%The experimental results confirm the effectiveness and viability of our methods in practical applications. This provides valuable insights for the design and optimization of RIS while emphasizing the potential value of RIS in wireless reverberant environments.

%These contributions collectively reveal the potential utility of RIS in strong reverberant environments, offering new insights and methods for research and development in the field of wireless communications. Our research findings provide robust support for addressing complex propagation environments and achieving reliable communication.
\end{abstract}

% Note that keywords are not normally used for peerreview papers.

% For peer review papers, you can put extra information on the cover
% page as needed:
% \ifCLASSOPTIONpeerreview
% \begin{center} \bfseries EDICS Category: 3-BBND \end{center}
% \fi
%
% For peerreview papers, this IEEEtran command inserts a page break and
% creates the second title. It will be ignored for other modes.
\IEEEpeerreviewmaketitle

\section{Introduction}

Deploying wireless communications in cavity-like environments, such as containers, naval vessels, aircraft, and spacecraft, presents substantial challenges. Typically, these environments have enclosed metallic surfaces with few apertures through which energy can leak out. Each wave traces a complex path involving multiple reflecting and scattering events over metallic boundaries, leading to extreme multipath propagation between the transmitter (Tx) and receiver (Rx) \cite{dupre2015wave}. The channel exhibits an unconventional appearance, to the extent that traditional spectrum recovery techniques (e.g., equalization) fail to restore the original signal \cite{PhysRevE.52.1146}. This results in a phenomenon of high signal strength but a high BER, as discussed in our previous research \cite{qiu2007time}.

A critical yet often underestimated aspect of enclosed cavities is the sensitivity of EM wave propagation to even minor boundary perturbations. In enclosed cavity-like environments, the free space continuum of solutions becomes a discrete set of eigenmodes of various eigenfrequencies due to scattering and interference. Importantly, these eigenmodes and associated eigenfrequencies are determined by the boundary conditions, intricately linked to the geometrical properties of the cavity. Despite several studies recognizing the channel's sensitivity to boundary perturbations \cite{dupre2015wave, gros2022multi, kaina2014shaping, frazier2022deep}, translating this insight into practical communication modulation schemes faces significant challenges. The most straightforward method involves the use of mechanical devices. Nonetheless, employing mechanical devices for wireless communication with high data rates seems infeasible.

In this paper, we propose a novel reconfigurable boundaries modulation scheme for wireless communications in cavities. The emergence of metasurface, or RIS, offers the possibility of electrically modifying the boundary of the cavity. Metasurfaces consist of regularly arranged EM units, typically sub-wavelength microstrip patches, printed on a dielectric substrate. Their reconfigurability is ensured by low-power tunable electronic circuits, e.g. positive intrinsic negative (PIN) diodes or varactors \cite{cui2014coding, xiong2023multirisaided}. By configuring the on/off states of the PIN diodes, we can control the propagation of reflected EM waves. Despite studies exploring modulation through the periodic variation of metasurfaces \cite{WTang2020, chen2022accurate, fara2021reconfigurable, zhang2021wireless}, it's crucial to note that this method is ineffective in cavity environments due to the high level of reverberation. We propose a shift in the paradigm, moving from the periodic modulation of RISs in the time domain to boundary modulation with RISs in cavities.

The contributions of this work are summarized as follows:   
\begin{itemize}
    \item We propose a reconfigurable boundary modulation framework for wireless communications in  enclosed cavities. Through a thorough analysis of eigenmodes, we demonstrate that even perturbations at the wavelength scale lead to substantial changes in eigenfrequencies. This forms the foundational principle for boundary modulations, which holds promise for addressing challenges inherent in cavity-like environments, particularly mitigating the high BER and tackling spectral selectivity issues.

    \item At the implementation level, we propose a novel RIS-generated equivalent PPM scheme, designed to accommodate the non-stationary nature of environments. We construct prototypes and conduct comprehensive testing to demonstrate its superiority in addressing the challenges posed by high reverberation.
\end{itemize}

The remainder of this paper is organized as follows. We start by examining wave propagation in reverberant cavities in Section II. Section III is devoted to the concept of reconfigurable boundary. In Section IV, we present the wireless communication schematic. Section V is dedicated to illustrating the obtained experimental results.

\section{Wave propagation in Reverberant Cavity}
For a dissipationless, electrically large cavity, the Helmholtz equation is derived from Maxwell's equations and associated boundary conditions. The eigensolutions of this equation provide eigenfrequencies and their coefficients. The transmission features are made up of these eigensolutions, which are determined by the placement of the Tx, Rx, and boundary. It is assumed that the walls of the $\boldsymbol{\Omega}$ fields with arbitrary boundary $\partial\boldsymbol{\Omega}$ act as an ideal conductor. The problem of modelling wave propagation through the use of Green’s functions can be simplified as a linear combination of the eigensolutions $\left\{\boldsymbol{\psi}_n(\boldsymbol{r})\right\}$ of Helmholtz equation, defined as
\begin{equation}
\small
\left\{
\begin{aligned}
    ( &\nabla^2+{k_n}^2 ) \boldsymbol{\psi}_n(\boldsymbol{r})=0, \quad \boldsymbol{r} \in \boldsymbol{\Omega} \\
    \partial &\boldsymbol{\Omega}=\partial \boldsymbol{\Omega}_{\mathrm{s}}+\partial \boldsymbol{\Omega}_d
\end{aligned}
\right.
,
\label{problem}
\end{equation}
where $k_n=\omega_n\sqrt{\mu\varepsilon}$, permittivity $\varepsilon$ and conductivity $\mu$ are both real, $\omega_n$ is the $n$-th eigenfrequency. $\partial\boldsymbol{\Omega}$ consists of the stationary part $\partial\boldsymbol{\Omega}_s$ and the dynamic part  $\partial\boldsymbol{\Omega}_d$.

However, in actual cavities, dissipation arises from wall losses, aperture leakage, object absorbability and other factors. As a result, the $n$-th eigensolution of (\ref{problem}) is attenuated by a particular decay rate $\tau_n$ due to the dissipation, resulting in the presence of the linewidth $\Delta \omega_n$ at the eigenfrequency $\omega_n$. 
\subsection{Eigenmodes}
The eigenmode $\boldsymbol{\psi}_n(\boldsymbol{r})$, also eigensolution, is the spectral response around the $n$-th eigenfrequency $\omega_n$. 
The electronic fields solution $E_n(t)$ corresponding to $\boldsymbol{\psi}_n(\boldsymbol{r})$, due to the energy dissipation in realistic reverberant environment, can be represented as:
\begin{equation}
\small
E_n(t)=E_{n 0} e^{-i \omega_n t} e^{-\frac{t}{2 \tau_n}}U(t),
\label{decay solution}
\end{equation}
where $U(t)$ is the unit step function, the $\tau_n$ is the exponential energy decay time and $E_{n0}$ is the initial electronic energy of $E_n(t)$. The absolute value of the Fourier transform of (\ref{decay solution}) is 
\begin{equation}
\small
\left|\tilde{E}_n(\omega)\right|=\frac{\left|E_{n 0}\right| \tau_n}{\pi} \frac{1}{\sqrt{1+\left[2 \tau_n\left(\omega-\omega_n\right)\right]^2}},
\label{decay solution fourier}
\end{equation}
which depicts the spectral response of the $n$-th mode. Equation (\ref{decay solution fourier}) suggests that a longer decay time $\tau_n$ results in a reduced linewidth $\Delta \omega_n$ for a particular mode of $E_n(t)$. Generally, the electric fields within interested range can be expressed as
\begin{equation}
\small
    \boldsymbol{E}(t) = \sum_{n=1}^{N_m}E_{n 0} e^{-i \omega_n t} e^{-\frac{t}{2 \tau_n}}U(t),
    \label{combinations}
\end{equation}
where $N_m$ is the average number of modes resonating within the interested bandwidth. For the channel of propagation, the combination of equation (\ref{combinations}) and (\ref{decay solution fourier}) results in greater spectral selectivity with a narrower linewidth $\Delta \omega_n$. 
\subsection{Boundary Perturbation}
\begin{figure}[ht]
\centering
\includegraphics[width=2.5in]{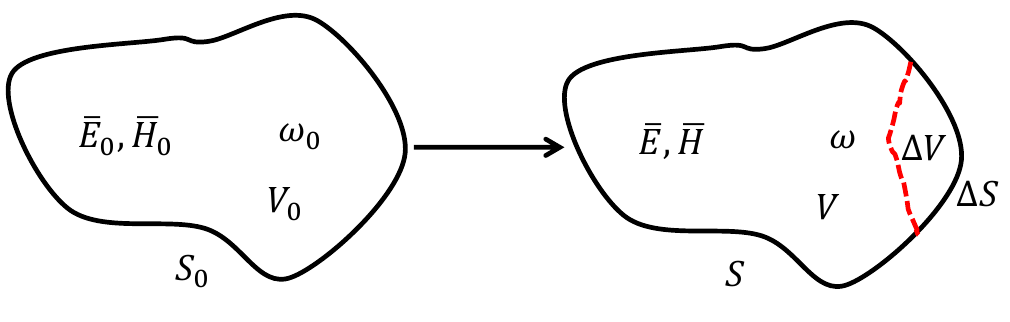}
\caption{A reverberant cavity perturbed by a change in shape of boundary.}
\label{shape_perturbation}
\end{figure}

A reverberant room is an environment with a high decay time. It have wave scattering properties that are quite sensitive to sub-wavelength perturbation, since the waves interact with the environment not once but countless times\cite{hill2009electromagnetic}. These perturbations are usually caused by boundary condition variations, which result in changes in eigenfrequencies. As shown in Fig.\ref{shape_perturbation}, we assume that the eigenfrequency of the $n$-th unperturbed mode is denoted as $\omega_n$, and the perturbed one as $\omega_n+\delta\omega$, then  
\begin{equation}
\small
    \delta\omega=\frac{-j \oint_{\Delta S} \bar{E}_0^* \times \bar{H} \cdot d \bar{s}}{\int_V\left(\epsilon \bar{E} \cdot \bar{E}_0^*+\mu \bar{H} \cdot \bar{H}_0^*\right) d v},
    \label{resonate_freq}
\end{equation}
where $\{\cdot\}^*$ denotes the conjugation. Equation (\ref{resonate_freq}) shows that the amount of change $\delta\omega$ in eigenfrequency is directly related to the perturbation on surface $\Delta S$. Proof of (\ref{resonate_freq}) in Appendix. 

Besides single mode, multiple modes exist within the operating bandwidth, which obey certain distribution, referred to as mode density. The eigenmodes counting function $N(\omega)$ can be represented as the cumulation of the modal spectral density $D(\omega)$ which is asymptotically equal to Weyl's law for a 3-dimensional and polarized EM field \cite{PhysRevE.52.1146}. Then 
\begin{equation}
\small
    N(\omega) \cong \int_0^{\omega} D(x) dx = \frac{\omega^3V}{3\pi^2c^3} = \frac{8\pi V}{3\lambda^3},
    \label{counting}
\end{equation}
where $V$ is the volume of the cavity, and $c$ is the vacuum light speed. The perturbation $\Delta V$ to the cavity's volume results in the mode number variation $\Delta N$. Set $\Delta N $ equal to $1$, and observe the volume ratio change,
\begin{equation}
\small
    \frac{\Delta V}{V} \cong \frac{3\lambda^3}{8\pi V}.
\end{equation}

Assuming that a cavity is a $2m\times2m\times4m$ cuboid ($V=16m^3$) and the excitation frequency is 3 GHz ($\lambda=0.1m$), then a small variation of boundary, i.e. $\Delta V \cong 1.2\times10^{-4}m^3$, will produce a perturbation to the field. 
\subsection{Eigenmodes-Based Reverberant Channel Model}
A reverberant field is an ensemble of equivalent, uncorrelated, and statistically identical field configurations introduced by different boundary conditions. While multipath propagation models, like the Saleh-Valenzuela model, exist, they do not offer a physical explanation of reverberation. 
%This section we propose a channel transform function based on the combination of modes.

According to (\ref{combinations}), the channel model can be separated to two part $h(t)=P(t)\omega(t)$, where the power delay profile envelope $P(t)$ and the statistical ensemble of modals $\omega(t)$. The former is expected to take the shape
\begin{equation}
\small
    P(t)=A_0e^{-t/2\tau}U(t),
\end{equation}
where we simplified the average decay time $\tau$ to time-invariant and frequency-invariant under the assumption that the ratio of bandwidth and canter frequency is not large. Usually, the average decay time can be measured by channel sounding. The statistical ensembles of modals $\omega(t)$ can be depicted as 
\begin{equation}
\small
    \omega(t)=\sum_{n=1}^{N_m}\alpha_n e^{j\varphi_n}e^{j\omega_nt},
\end{equation}
where we can obtain three statistical ensembles $\left\{ \alpha_n\right\}$,$\left\{ \varphi_n\right\}$ and $\left\{ \omega_n\right\}$, which respectively denotes the modal coefficients, random phase-shift angles and random resonant frequencies. Once the fields becomes reverberant, the field has the statistical characteristics of spatial uniformity and isotropy, which are sort of benefit to our modeling\cite{6178273}. We can rewrite the channel in angular domain:
\begin{equation}
\small
    H(\omega; \left\{ \alpha_n\right\},\left\{ \varphi_n\right\},\left\{ \omega_n\right\})=A_0\sum^{N_m}_{n=1}\frac{\alpha_n e^{j\varphi_n}}{j(\omega_n-\omega)+\frac{1}{2\tau}}.
    \label{channel augular domain}
\end{equation}

Specifically, the $\left\{ \alpha_n\right\}$ is Gaussian distributed with an expectation related to cavity volume and quality factor, due to the central limit theorem. The statistical isotropy leads to $\left\{ \varphi_n\right\}\sim\mathcal{U}(0,2\pi)$, i.e., uniformly distributed over all possible angles. Moreover, the realization of the ensemble $\left\{ \omega_n\right\}$ are sensitive to the boundary conditions although the ensemble has also some characteristics in terms of number and inter-distance, where the number follows (\ref{counting}) and the eigenfrequency spacing is close to Wigner distribution \cite{PhysRevE.52.1146}\cite{6178273}. As shown in Fig.\ref{two_PSD}, the two codebooks correspond two realizations of above three ensembles of the channel transformer function in (\ref{channel augular domain}).

\section{Reconfigurable Boundary}
As discussed in Section II, the wave field within a reverberant cavity is statically fixed by its boundary. These boundaries are typically modified with mechanical parts such as rotating paddles in the domain of EM compatibility. 

\begin{figure}[ht]
\centering
\includegraphics[width=3.5in]{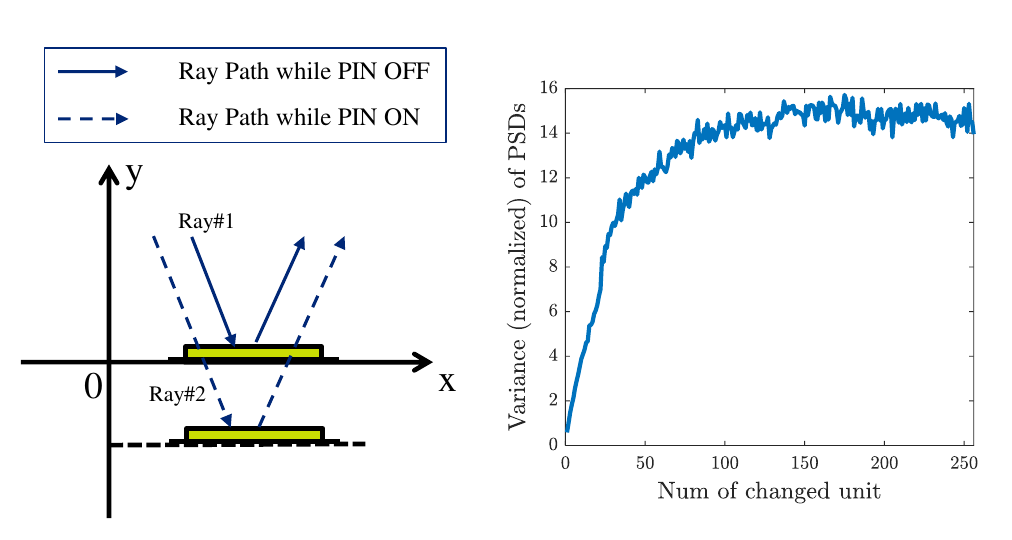}
\caption{(1) Left: Equivalent physical position of unit when the PIN ON/OFF; (2) Right: the variance of normalized power spectral density (PSD) when increasing the number of changed units.
}
\label{equivalence}
\end{figure}

The units of RIS is usually made of PIN diode and designed to simply produce the phase difference\cite{xiong2023design}. The variable phase state can be controlled by applying the state "0"  or "1" to the PIN diodes. Its reconfigurability gives cavities the capacity of arbitrarily changing their boundaries. The phase shift of unit can be equivalent to the vertical displacement as impinging EM wave is treated from the perspective of ray model as shown in the left portion of Fig.\ref{equivalence}. The propagation distance of the ray, which is emitted from the source and reflected by the unit then sent to the sink, will increase around $\Delta d=\frac{\Delta \phi}{2\pi}\lambda$ as the state switches from OFF to ON. For the wave with center frequency at 3GHz ($\lambda=0.1m$), the phase shift $\pi$ of one unit of RIS produces an equivalent perturbation of $5\times 10^{-4}m^3$, which is larger than the critical volume in Section II B.

We validates this concept of the equivalent perturbation in our testbed. As depicted in the right portion of Fig.\ref{equivalence}, the perturbation of resonant frequencies is directly proportional to number of units.
In conjunction with (\ref{problem}), the dynamic boundary condition $\partial \boldsymbol{\Omega}_d(t)$ in the cavity is expressed as
\begin{equation}
\small
\partial \boldsymbol{\Omega}_d(t)=\partial \boldsymbol{\Omega}_{R I S}\left(\left\{\boldsymbol{\Phi}_n\right\}\right)+\partial \boldsymbol{\Omega}_{\text {scatter }}(t),
\end{equation}
where $\partial \boldsymbol{\Omega}_{\text {scatter }}(t)$ and $\partial \boldsymbol{\Omega}_{R I S}\left(\left\{\boldsymbol{\Phi}_n\right\}\right)$ respectively correspond to boundary of moving scattering objects and different codebooks, $\left\{\boldsymbol{\Phi}_n\right\}$ denotes the codebooks sequences. Besides the $\partial \boldsymbol{\Omega}_{\text {scatter }}(t)$, the
\begin{equation}
\small
\partial \boldsymbol{\Omega}_{R I S}(t)=\sum_{n=1}^{\infty} \partial \boldsymbol{\Omega}_{R I S}\left(\boldsymbol{\Phi}_n\right)\left[U\left(t-t_{n}\right)-U\left(t-t_{n+1}\right)\right],
\label{ris boundary}
\end{equation}
where $t_n$ is the switching moment of $n$-th codebook, $\partial \boldsymbol{\Omega}_{R I S}\left(\boldsymbol{\Phi}_n\right)$ represents the boundary related to $\boldsymbol{\Phi}_n$.

\begin{figure}[ht]
\centering
\includegraphics[width=2.5in]{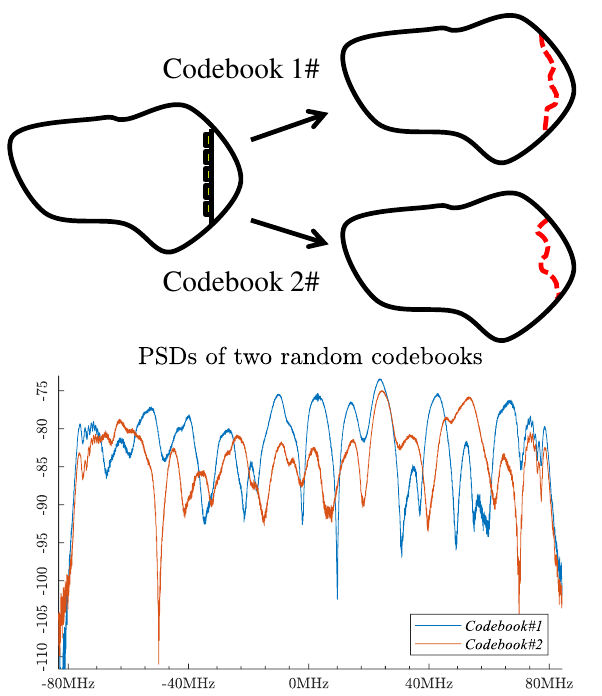}
\caption{Front: Two random codebooks be equivalent to two different perturbations to the cavity; Below:  PSD received by a stationary omnidirectional antenna when the Tx is transmitting a frequency-flat signal.
}
\label{two_PSD}
\end{figure}

Based on the above idea, we conducted a simple experimental verification where the specific settings details will be described in the section of system implement later. As shown in Fig.\ref{two_PSD}, the received PSD has been strongly different due to the different cookbooks applied to the RIS in cavity when the Tx consistently transmits a frequency-flat signal.

\section{A Reconfigurable Boundary Modulation Approach}
\subsection{System Formulation}
In our envisioned communication system, the information is conveyed through the manipulation of EM fields, achieved by controlling the reconfigurable boundary. According to the deterministic theory \cite{hill2009electromagnetic}, a certain boundary conditions give a realization of the three ensembles in (\ref{channel augular domain}), which denotes as $\partial \boldsymbol{\Omega}_{d} \longleftrightarrow (\left\{ \alpha_n\right\},\left\{ \varphi_n\right\},\left\{ \omega_n\right\})$.The system can be formulated as fellow:
\begin{equation}
\small
    Y(\omega,t)=H(\omega;\partial\boldsymbol{\Omega}_d(t))X(\omega,t).
    \label{continuous system formulation}
\end{equation}

In practical communication systems, signals are typically transmitted in a frame-based format. Variations occurring between frames are categorized as time-varying, while variations within frames are attributed to propagation delays. Here we discrete the time $t$ into frame indices $n\Delta t$:
\begin{equation}
\small
    Y[\omega,n]=H[\omega;\partial\boldsymbol{\Omega}_d[n]]X[\omega,n],\quad n=1,2,\dots,N
    \label{discrete system formulation}
\end{equation}
\begin{figure*}[tbp]
	\centering
	\subfloat{\includegraphics[width=3in]{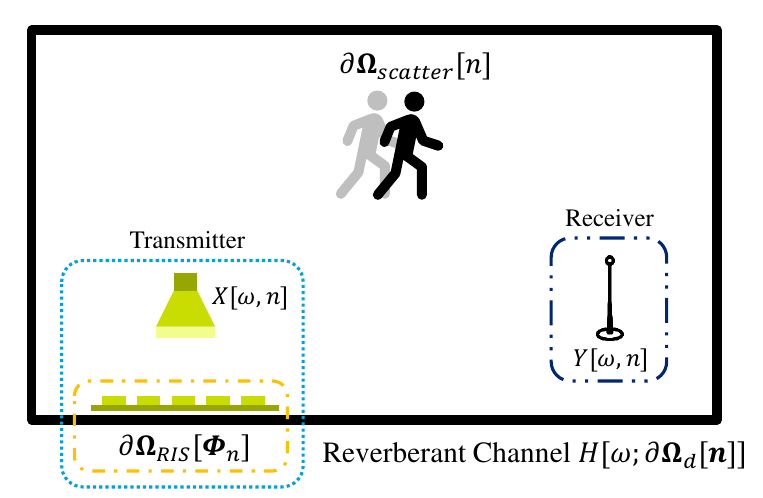}}\quad
	\subfloat{\includegraphics[width=3in]{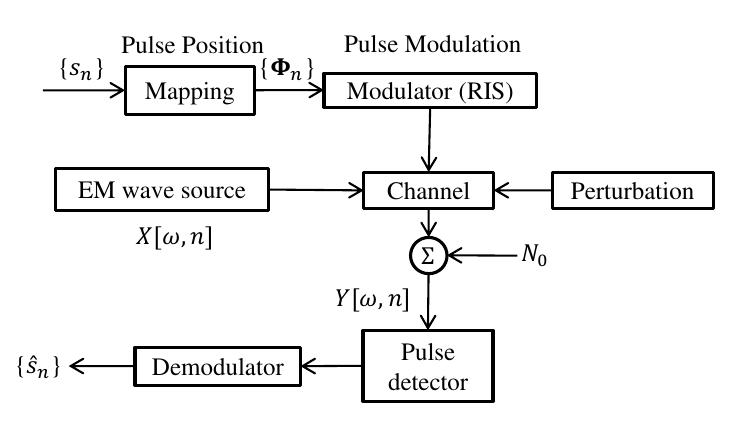}}\\	
	\caption{(a) Left: The scene schematic of reconfigurable boundary modulation in cavity. (b) Right: The RIS-generated equivalent pulse position modulation scheme.}
 \label{PPM}
\end{figure*}
where $\Delta t$ denotes the frame period and $X[\omega,n]$ represents a readily generated source of consistent broadband EM wave signals, such as broadband white noise or Gaussian pulses, which are continuously irradiated onto the RIS satisfying $X[\omega,i]=X[\omega,j],\forall{i,j}$.  As shown in Fig.\ref{PPM} (a), the information stream does not through the EM wave source. This is intended to ensure that any variations observed at Rx are solely attributable to the $\partial\boldsymbol{\Omega}_d[n]$:
\begin{equation}
\small
    \partial\boldsymbol{\Omega}_d[n]=\partial\boldsymbol{\Omega}_{scatter}[n]+\partial\boldsymbol{\Omega}_{RIS}[\boldsymbol{\Phi}_n].
    \label{discrete boundary}
\end{equation}
\subsection{Wave Source}
For the EM source, we are concerned with two aspects: (i) the waveform of each frame; (ii) the direction of the EM radiation. Following the discussion of eigenmodes in Section II, our radiated signal must contain a sufficient number of eigenmodes to better distinguish the changes in PSD due to different codebooks. So it's appropriate to use a broadband signal as the waveform of each frame, such as a Gaussian pulse signal, a linear frequency modulation (LFM) signal, etc.

In order to maximise the impact of the reconfigurable boundary, we must ensure that a significant proportion of $\partial \boldsymbol{\Omega}_{RIS}$ is included in $\partial \boldsymbol{\Omega}$. Fully decorating all walls of the cavity with RIS poses certain difficulties, so instead we opt to ensure that initial wave reflection occurs at the surface of the RIS, as illustrated in Fig.\ref{PPM} (a). This approach allows for the RIS to have the greatest effect on the field within the cavity.
\subsection{RIS-generated Equivalent PPM}
The motion of both the scatterer and the Rx can be viewed as modifications in boundary conditions $\partial\boldsymbol{\Omega}_{scatter}$. As demonstrated in Section II.B, the cavity displays remarkable sensitivity to changes in internal boundary conditions, which is supported by the presentation of equations and examples. The variation of boundary conditions occurs on both terms in (\ref{discrete boundary}), where both the differences $\partial \boldsymbol{\Omega}_{scatter}[n]-\partial \boldsymbol{\Omega}_{scatter}[n-1]$ and $\partial \boldsymbol{\Omega}_{RIS}[\boldsymbol{\Phi}_n]-\partial \boldsymbol{\Omega}_{RIS}[\boldsymbol{\Phi}_{n-1}]$ are non-zero. Distinguishing between the respective variations introduced by these two terms is the objective of the detection mission.

The motion of objects in the physical world results in the continuous variation of boundary over time within a enclosed environment. By contrast, the switching of the RIS codebook introduces a discontinuity, in which the boundary undergo abrupt changes over time, where
\begin{equation}
\small
\frac{\partial^2 \boldsymbol{\Omega}_{R I S}\left(\left\{\boldsymbol{\Phi}_n\right\}\right)}{\partial t}=\sum^{\infty}_{n=2}\left[\partial \boldsymbol{\Omega}_{R I S}\left(\boldsymbol{\Phi}_n\right)-\partial \boldsymbol{\Omega}_{R I S}\left(\boldsymbol{\Phi}_{n-1}\right)\right] \delta\left(t_n\right)
\label{pulse}
\end{equation}
and $\forall t\in \mathbb{R}$, $\exists M$, $\arrowvert \frac{\partial^2 \boldsymbol{\Omega}_{\text {scatter }}(t)}{\partial t}\arrowvert < M$.

The RIS codebook switching may be likened to a pulse in (\ref{pulse}), and discerning whether variance in the received signal is due to scatterer movement or the codebook requires an operation akin to differential detection from one frame to the next. Pulse modulation techniques, such as pulse position modulation (PPM), are used to convey information. As depicted in Fig.\ref{PPM} (b), we present an innovative communication framework to tackle the vulnerability to moving objects inside the cavity. 
\subsection{Pulse Detector}
The key aim of this methodology is to distinguish whether the codebook has undergone a switch or not, with the aid of the pulse detector (PD), accomplishing this aim by examining the discrepancies between frames (specifically $Y[\omega,n-1]$ and $Y[\omega,n]$). Ultimately, the primary objective is to determine a suitable distance function, which is referred to as $d(\cdot)$, and an appropriate threshold value, referred to as $\eta$, that fulfil the subsequent guidelines:
\begin{equation}
\small
    PD[n]=\left\{
\begin{aligned}
    1, \quad d(Y[\omega,n-1],Y[\omega,n]) \geq \eta \\
    0, \quad d(Y[\omega,n-1],Y[\omega,n]) < \eta
\end{aligned}
\right.
.
\label{pd}
\end{equation}

\begin{figure}[ht]
  \centering
\includegraphics[width=2in]{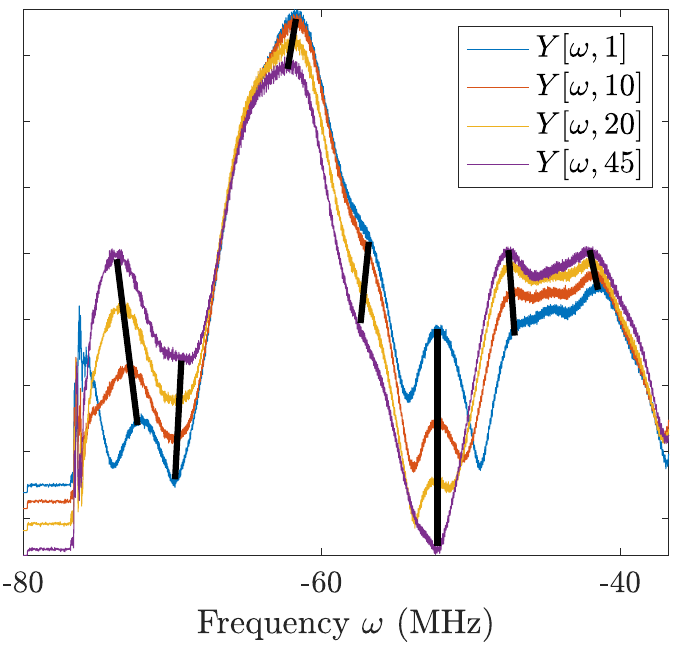}
\caption{For visual clarity, we plot the partial spectrum of four frames respectively received at $n=1,10,20,45$, where $\Delta t=50\mu s$, in the scenario that the boundary variation is only the scatter object's movement. The black lines represent the trajectories of certain eigenmodes}
\label{pattern sequences}
\end{figure}

To examine the temporal characteristics of spectral sequences when objects are moving (the Rx movement is equivalent to object movement), we analyse the series of frames captured by the testbed shown in Fig. \ref{systemimplement}. As illustrated in Fig.\ref{pattern sequences}, the eigenmode trajectories of consecutive frames are not parallel, hence, Euclidean distance or auto-correlation cannot extract their similarity. Instead, alignment-based metrics like dynamic time warping (DTW) \cite{berndt1994using} should be employed.
Based on the temporal characteristics, it is observed that two consecutive frames exhibit high similarity when the time interval $\Delta t$ is sufficiently small, despite the scatterers' movement. We can conclude that for RIS-generated equivalent PPM schemes, a shorter frame period $\Delta t$ results in a higher symbol rate and greater resistance to perturbation. 
\section{Prototype Implementation}
\begin{figure}[ht]
\centering
\includegraphics[width=3.3in]{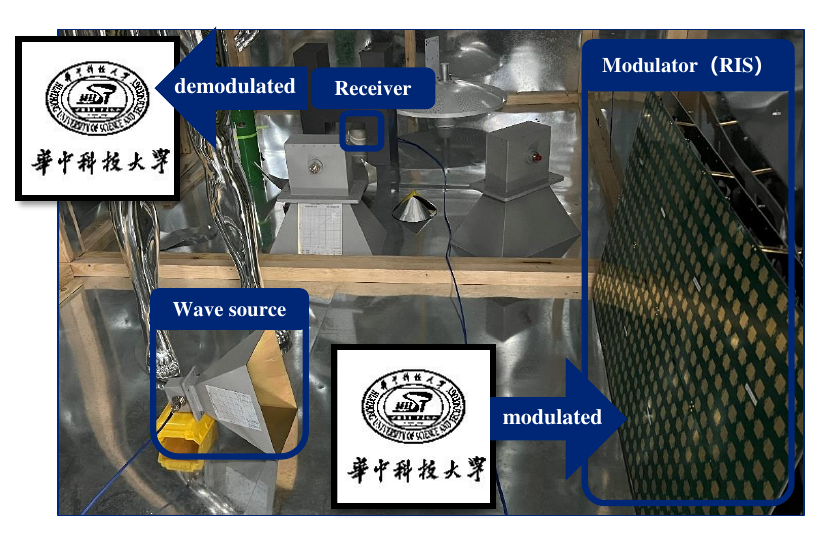}
\caption{The testbed of the RIS-modulated system, in which a picture has been transmitted from RIS to Rx by the RIS-generated equivalent pulse position modulation.}
\label{systemimplement}
\end{figure}

To validate the proposed modulation schemes devised for reverberant environments, we constructed a metallic cavity using iron and wooden bars as Fig.\ref{systemimplement} ($2m\times 4m\times 2m$). Our testbed consists of NI Universal Software Radio Peripheral (USRP) X310, which generates the frame-based broadband signals and perceives the EM fields. The RIS functions as a modulator consisting of 512 units whose center operating frequency is 3.3 GHz. As described in Section IV.B, the horn antenna is positioned as the Fig.\ref{systemimplement}. We choose the LFM signal as the EM wave source which has 160MHz bandwidth and the same center frequency as the RIS's. Each LFM frame comprises 8192 samples at a sampling rate of 160MHz, resulting in roughly 50 $\mu s$ per frame slot. 
%The minimum frame period $\Delta t$ of USRP is 20ms due to the CPU performance. 

\begin{figure}[ht]
    \centering
    \includegraphics[width=3.5in]{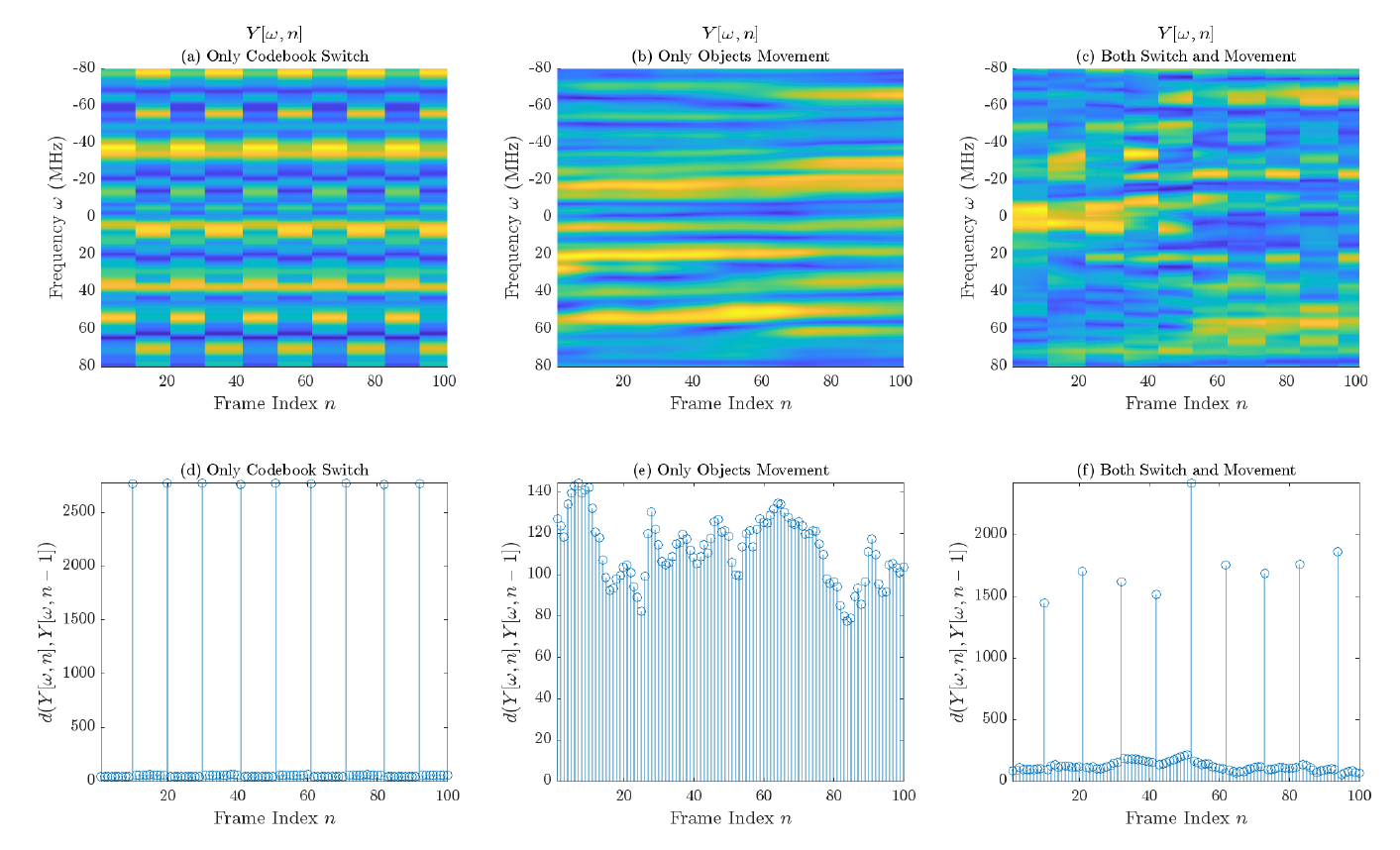}
    \caption{The received frame sequences $Y[\omega,n]$ and the PD's output. (1) First Row: The three depicted graphs illustrate spectral sequences for three distinct scenarios. The horizontal axis denotes the frame index (total 100 frames), while the vertical axis corresponds to the baseband spectrum (from -80MHz to 80MHz); (2) Second Row: The three figures show the results of DTW algorithm acting on the corresponding spectral sequences above, respectively. }
    \label{three scence}
\end{figure}

We collected temporal sequences for spectral analysis in three different scenarios, as shown in Fig.\ref{three scence}, to verify pulse detection outcomes. We can clearly observe that the spectral sequence regularly changes between the two states when only the codebook switches. The second figure validate that the temporal continuity of the boundary results in the continuity of field's variation over time. This continuity can be broke by the reconfigurable boundaries using RIS as shown in the third figure. The corresponding similarity measurements, which is applied by the DTW algorithm, between the consecutive frames are plotted below. Despite the movement of the scatterers, we can still precisely determine the moment of codebook switching, which is crucial for us to transmit information using RIS-generated equivalent PPM.

We used the gap between equivalent pulses to decide which symbols to transmit, which to some extent overcomes the problem of inaccuracies of switches timing due to instability in the control flow. We transmitted a picture by the RIS-generated Equivalent PPM and received it with zero BER. The data rate of the prototype system reaches around 2 Mbps when ignoring the overhead of the synchronization and redundant frames. And we compared the performances of the proposed modulation signal and the equalized orthogonal frequency division multiplexing (OFDM) signal under the same dynamic scenario in the cavity. As shown in the Table.\ref{tab}, the proposed scheme exhibits a strong capability to accurately demodulate signals in environments with reverberations and moving objects even without any channel or error correcting coding.
\begin{table}[h!] 
    \caption{The BER when employing two modulation schemes without any channel or error correcting coding. }
    \centering
    \resizebox{0.48\textwidth}{!}
    {
    \begin{tabular}{cccccc}                
    \toprule
    \multirow{2}{*}{\textbf{Modulation}}  & \multicolumn{3}{c}{\textbf{RIS-generated Equivalent PPM}} & \multicolumn{2}{c}{\textbf{OFDM}} \\
    \cmidrule(lr){2-4}\cmidrule(lr){5-6}
    & stationary & walking & running & equalized \\
    \midrule
    \textbf{BER (fixed SNR)}\quad & 0.0000 & 0.0000 & 0.0732 & 0.4969 \\
    \bottomrule
    \end{tabular}
    }
    \label{tab}
\end{table}
\section{Conclusion}
In this paper, we have introduced a communication paradigm with innovative boundary modulation schemes using RIS to address scenarios in reverberant environments. Based on this paradigm, we proposed a RIS-generated equivalent PPM to deal with the wireless communication's issues in reverberant environment. Future research endeavors will explore more abstract modulation schemes by harnessing new degrees of freedom and will center on multi-user scenarios.

\appendices
\section{Proof of Equation (\ref{resonate_freq})}
As shown in fig.\ref{shape_perturbation}, let $\bar{E}_0, \bar{H}_0, \omega_0$ be the fields and resonant frequency of the original cavity and let $\bar{E}, \bar{H}, \omega$ be the fields and resonant frequency of the perturbed cavity with fixed permittivity $\epsilon$ and permeability $\mu$. Making use of Maxwell's curl equations, we obtain
\begin{subequations}
\small
\begin{align}
& \nabla \times \bar{E}_0=-j \omega_0 \mu \bar{H}_0,\label{a}\\&\nabla \times \bar{H}_0=j \omega_0 \epsilon \bar{E}_0,\label{b}\\
& \nabla \times \bar{E}=-j \omega\mu \bar{H},\label{c} \\&\nabla \times \bar{H}=j \omega\epsilon \bar{E}.\label{d}
\end{align}
\end{subequations}

Now multiply the conjugate of (\ref{a}) by $\bar{H}$, multiply (\ref{d}) by $\bar{E}^*_0$, multiply (\ref{b}) by $\bar{E}$ and multiply (\ref{c}) by $\bar{H}^*_0$. Then subtracting each other and applying the vector identity $\nabla \cdot\left(A \times B\right)=B\cdot\nabla\times A-A\cdot\nabla\times B$ then gives
\begin{subequations}
\small
\begin{align}
& \nabla \cdot\left(\bar{E}_0^* \times \bar{H}\right)=j \omega_0 \mu \bar{H} \cdot \bar{H}_0^*-j \omega \epsilon \bar{E}_0^* \cdot \bar{E}, \label{2a} \\
& \nabla \cdot\left(\bar{E} \times \bar{H}_0^*\right)=-j \omega \mu \bar{H}_0^* \cdot \bar{H}+j \omega_0 \epsilon \bar{E} \cdot \bar{E}_0^*. \label{2b}
\end{align}
\end{subequations}
Now add (\ref{2a}) and (\ref{2b}), integrate over the volume $V$, and use the divergence theorem  to obtain 
\begin{equation}
\small
\begin{aligned}
& \int_V \nabla \cdot\left(\hat{E} \times \bar{H}_0^*+\bar{E}_0^* \times \bar{H}\right) d v\\=&\oint_S\left(\bar{E} \times \bar{H}_0^*+\bar{E}_0^* \times \bar{H}\right) \cdot d \bar{s} \\
=&\oint_S \bar{E}_0^* \times \bar{H} \cdot d \bar{s}\\=&-j\left(\omega-\omega_0\right) \int_V\left(\epsilon \bar{E} \cdot \bar{E}_0^*+\mu \bar{H} \cdot \bar{H}_0^*\right) d v,\label{integral}
\end{aligned}
\end{equation}
since $\hat{n}\times \bar{E}=0$ on $S$. Since the perturbed surface $S=S_0-\Delta S$, we can write
\begin{equation}
\small
\begin{aligned}
        \oint_S \bar{E}_0^* \times \bar{H} \cdot d \bar{s} =&\oint_{S_0} \bar{E}_0^* \times \bar{H} \cdot d \bar{s}-\oint_{\Delta S} \bar{E}_0^* \times \bar{H} \cdot d \bar{s}\\ =&-\oint_{\Delta S} \bar{E}_0^* \times \bar{H} \cdot d s,
\end{aligned}
\end{equation}
because $\hat{n}\times \bar{E}_0=0$ on $S_0$. Using this result in (\ref{integral}) gives equation (\ref{resonate_freq}).

\ifCLASSOPTIONcaptionsoff
  \newpage
\fi
%\printbibliography %Prints bibliography
% Generated by IEEEtran.bst, version: 1.14 (2015/08/26)

\bibliographystyle{IEEEtran}
\bibliography{ref}
\end{document}